# Slowing Plants, Slowing Home

Proposing A Plant-Decentred Perspective to Human-Plant Interaction

XINQUAN WEN

School of Design, The Hong Kong Polytechnic University, xinquan.wen@outlook.com

YIYING WU

School of Design, The Hong Kong Polytechnic University, bow.yiying.wu@gmail.com

The Anthropocene is causing a global crisis in recent decades. Facing this challenge, increasing attempts are being made to explore the more-than-human-centred perspective in HCI and design. Our research sets out to explore the ways of experiencing and interacting with plants with a case study on the slowness of plants. Utilising existing time-lapse technology, we investigate the role of IoT technologies in associating biological slowness with the networked technological environment of the home. In the experiment, we chose the humidity level of the environment as the variable to synchronise the movement of smart curtains and plants. We propose a relationship-centred strategy that uses an inclusive feature of a microclimate, like humidity, instead of the plant itself, for human-plant interaction. Furthermore, it indicates a 'plant-decentred' perspective to spark critical reflection on the taken-for-granted perception of isolating a person or a plant as an individual entity.

## 1 INTRODUCTION

The Anthropocene is causing a global crisis in recent decades, such as climate change. To face this challenge, the HCI community is shifting from the human-centred design paradigm to the posthuman agenda. This agenda questions the central position of humans and human values, and non-human agents, such as plants [11, 18], and technological devices [9], are understood no longer as overlooked objects but as active agents. Based on the more-than-human-centred perspective, previous works investigate multispecies and their interdependencies [13, 14], co-design and cohabitation with nature [15].

Following the non-anthropocentric shift, this study is interested in exploring different modes of knowing and experiencing between humans and plants. With interactive technologies, humans can interact or communicate with plants [2]. However, many interactive prototypes are designed to use plants as input or output devices or biosensors [2, 19, 21]. This approach sees one or part of the plant, such as a stem, as an individual entity. However, unlike humans, plants organise themselves differently and live in a more decentralised and non-hierarchical structure [16]. Thus, we attempt to see through the eyes of plants and then experiment with new ways of interacting with and experiencing plants. Regarding experiencing plants, our study focuses on the aspect of timescale, which is the slowness of plants. In biology, 'slow' is a prominent feature of plants whose metabolism and growth are slower than most animals [12]. Built on a widely used slow technology, time-lapse, our experiment intends to use IoT technologies to associate biological slowness with the networked technological environment.

In this short paper, we report a case study in which we built an IoT-based interactive installation called 'Sloooooooow', which uses the letter o ten times. By giving this obscure name, we express our wish for multiple options for humans to experience slowness, including at the moment of reading this name. With this interactive installation, we synchronise the movement of an IoT-based Curtain System with the movement of plants. And as a return, the movement of the curtain would influence the photosynthesis of the plants. To connect the two, we utilise the environment's humidity level as the attribute. Moreover, our experiment speaks to the slow technology investigated in the HCI community [10, 17]. However, expanding the realm of the interaction between humans and machines, we intend to explore the slowness in a non-anthropocentric way, involving humans, plants and technological devices.

This experiment demonstrates a novel way of experiencing and interacting with plants through the lens of IoT technology and how such technology-mediated experience affects humans and plants' daily life. We propose a valuable strategy for human-plant interaction, which is a relationship-centred strategy that focuses on the humidity level of a microclimate. Based on this strategy, we suggest a perspective called 'plant-decentred' as the explorative attempt. By giving this provocative name, we intend to spark critical reflection on the taken-for-granted perception of isolating a person or a plant as an individual entity.

## 2  RELATED WORK

Except for a few noticeable rapid plant movements like Mimosa and the Venus Flytrap, most plants are slow and operate on a timescale beyond human senses [5]. To enable humans to see or sense the timescale of plants, researchers often use the method of anthropomorphising plants, such as utilising capacitance changes or other instant data of plants to allow them to enter the human timescale. In this part, we review two sets of strategies of shifting models of experiences to enhance human-plant interaction. One is to train plants to learn to communicate spontaneously in a way humans understand. The other is to facilitate humans' ability to sense the timescale of plants with methods such as time-lapse.

The first set often refers to the strategy of humanising nature. Researchers translate the signals of plants into human feelings. The technologies range from hearing plants through sonification [1] to feeling plants through soft robotics [6, 20]. More specifically, for instance, Christiansen et al. [2020] added a digital layer over plants that leads to sensory extension and technology-aided synesthesia. In line with Christiansen's team, we believe that a posthuman entity is a composition of multiple intertwined actors. In addition, interactive devices allow plants to speak. For instance, electrodes can measure the shifts in plant electrical signals when plants respond to environmental stimuli [8]. The instant and verifiable voltage changes serve as the input of plants in several plant-human interactive prototypes, e.g., pheB [20] and Homo Viridis [6].

The second part refers to the techniques that invite humans to enter the timescale of plants. The most basic technology is time-lapse photography, which can compress hours into a few seconds. For instance, from the artwork 'The Birth of a Flower' [23] to the BBC documentary 'The Green Planet' [25], the condensing of time demonstrates how leaves bud and unfold, flowers bloom and wither, and stems whirl and grow [4]. Besides time-lapse, the project 'The Idea of a Tree' [24] presents an alternative to document natural input beyond the human timescale. A solar-powered installation produces one object daily, whose thickness and scale depend on the sun's intensity. The project reveals how plants, as immobile beings, have a different relationship to locality than humans do. Within this project, the complexity of locality, including each cloud and shadow, can present itself through the variable, which is the intensity of sunlight.

## 3  THE EXPERIMENT CASE

This part describes the case of the interactive installation named 'Slooooooooooow' and the experiment process conducted in one bedroom for about ten days.

## 3.1 The 'Slooooooooow' Installation

The 'Slooooooooow' installation is an interactive system that can synchronise the movement of the curtain with the growth and movement of plants, and the process of movement was recorded as a time-lapse. 'Slooooooooow' challenges the human-centredness of IoT home automation by giving plants the majority of control over the opening of a curtain. In return, the photosynthesis of the plants is influenced by the curtain's opening. Therefore, the input-output of the interactive system can be completed without the involvement of humans.

To achieve that, it connected the curtain's opening to the microclimate's humidity level, focusing on the transpiration of plants. Every 25 seconds, the system read the microclimate's humidity level, analysing whether the transpiration of plants dominates the change and deciding whether to open or close the curtain by 4%. Every 30 seconds, a photo was taken for time-lapse. In the actual building process, we combined an existing smart curtain and its app from the company 'Xiaomi' and its IoT domestic ecosystem called 'Mijia' with Arduino sending controlling instructions from the speech synthesis module of Arduino to the Xiaomi smart speaker. 'Slooooooooow' operated on the sill of a bedroom, where the window faced south. Moreover, there was little human activity during the daytime when the system ran. The well-being of the plants was considered during their selection and the experiment process. Monstera Deliciosa, Clivia Miniata, and Jasmine Sambac, all of which could survive and thrive under indirect light. The curtain's opening was constrained between 50% and 90%, to ensure adequate daylight levels were maintained.

The hardware part of the 'Slooooooooow' installation was composed of the aforementioned Xiaomi curtain, Xiaomi smart speaker, Arduino, speech synthesis module, and environment sensor module (Figure 1). Arduino checked the changes in humidity level every 25 seconds, and only change within 0.1-0.3% could trigger the curtain movement, which ensured the transpiration was the main cause of the curtain movement and filtered out most environmental errors. When the humidity level increased between 0.1% and 0.3%, the curtain would close by 4%, and vice versa. The Xiaomi curtain could be controlled from 0-100% in 1% increments. However, due to the storage limitations of Arduino, only ten instructions were stored within the range of 50 -90% in 4% increments.

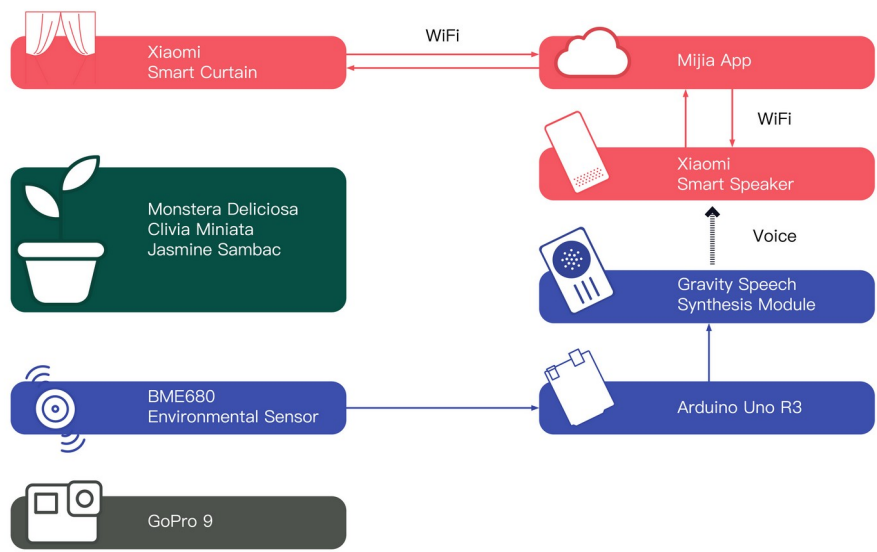

Figure 1: The components of the interactive installation 'Slooooooooow'

### 3.2 The Experiment: Slow Plants, Slow Home

This part outlines the integration of the human resident during the experiment, who co-lived with three plants with the 'Slooooooow' installation (Figure 2), who is also the first author and the designer who built the installation. The experiment was conducted in his bedroom over ten days. This experiment was carried out during the Shanghai lockdown. Therefore, we could not find other people to participate in the interactive installation. During the ten days, the resident took field notes on his experience and observations related to his bedroom and the installation.

On the site, the movement of the 'Slooooooow' technological devices was explicit for the resident. He reported that he could feel the 4% curtain movement accompanied by the sound of the motor and trembling of the curtain, following the conversations between Arduino and the Xiaomi smart speaker. Theoretically, he knew that the photo-synthesis process of plants was actually influenced by the sunlight received, which was controlled by the curtain. However, the atmospheric changes due to transpiration and photosynthesis were beyond his sensory capabilities. Moreover, he was not sensitive to the changes in atmospheric humidity when they were within the human comfort spectrum of between 30% and 50% (Mayo clinic). Nevertheless, every instruction sent by Arduino reminded him that the humidity level had changed between 0.1-0.3% in the past 30 seconds. He tried to explain the tiny temporal fluctuations: were they because of the transpiration of the plants? Or his presence? Or the wind? Furthermore, he knew that Swiss Cheese Plants' comfort humidity level (of over 50%) was higher than humans. Therefore, he thought the variation of humidity levels of microclimate could reflect a continuous cross-species negotiation.

Apart from the on-site experience, he watched the time-lapse of the room, including the movement of plants and curtains. They made him realise that the movements of plants occur at various timescales and humans are sensitive to certain speeds of movement. When he watched the time-lapse at 1x speed, he saw the curtain swaying; when watching the time-lapse at 3x speed, he saw the Bush Lily flaps its leaves; and when watching the time-lapse at 5x speed, he saw the stems of the Swiss Cheese Plant nod.

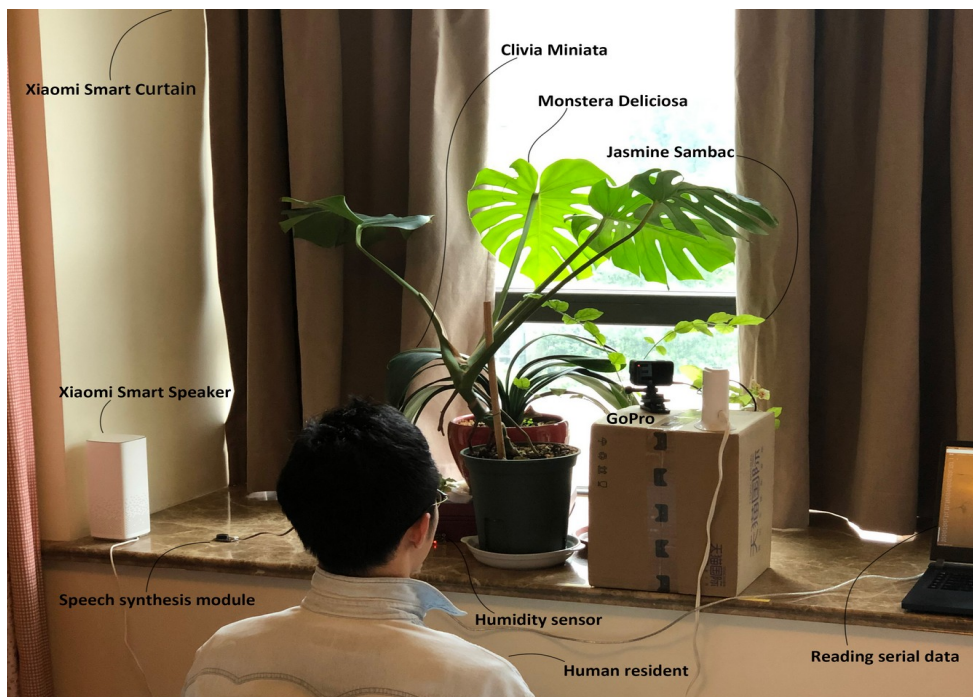

Figure 2: The IoT home with three plants and smart curtains

## 4 FINDINGS

After presenting the experiment of 'Sloooooooooow', we discuss the learnings from this experiment on how to design interactive systems to experience plants. The first finding discusses the relationship between the human's capability to sense the richness of slowness and the technology utilised. And secondly, we propose a plant-decentred perspective as an explorative attempt for human-plant interaction.

### 4.1 Slowness Embedded Technology

Slowness is relative and abundant, some of which can be sensed by humans and some not. And time-lapse technology can reveal the broad and rich spectrum of slowness beyond the limited human senses. In our experiment, the curtain, the leaves of the Bush Lily, and the stems of the Swiss Cheese Plant provided references for humans to enter beyond-human sensory timescales. The multiple timescales unfolded by time-lapse reminded the human resident of how he was confined to an accelerating life pace driven by digital technologies and neglected the rhythms of nature on his sill. In comparison, indigenous people know how to work with nature at the timescale of plants according to their traditional ecological knowledge [4]. For instance, the Khasis people in Meghalaya know how to build living root bridges and root ladders in cooperation with the biological growth of rubber fig trees.

Hallnäs & Redström's groundbreaking work [2001] has shed light on the importance of slow technology. However, their design agenda regarding slow technology is human-centred. Further studies of slow technology include designing for reflection, mental rest, and multi-lifespan [7, 10]. Whilst in our work, the inclusion of rhythms in plants' movements expands the human-centred design philosophy for slow technology. The plant kingdom provides a coordinate system within the unlimited universe of slowness beyond or on the edge of our senses, which opens the design space for communicable slowness. For instance, do you want your breathing light to breathe at the pace of the stems of a spring Bush Lily or an autumn Swiss Cheese Plant? Slowing down with plants is reflective and calming due to its non-human-centred and natural denotations, which aligns with the current design philosophy for slow technology.

Moreover, the ability to travel across timescales relies on and is confined to the utilised technology. In our experiment, the time-lapse was shot by GoPro 9 and edited in Adobe Premiere. The time-lapse mode of GoPro 9 provided 11 interval options from half a second to one hour under video format. The choices are abundant for daily usage, nevertheless, the limitation of its capabilities to travel across timescales is embedded.

### 4.2 A Plant-Decentred Perspective: Subtlety and Ambiguity of Locality

Next, we propose a 'plant-decentred perspective' in designing for human-plant interaction. This perspective, first of all, is based on a central concept of the sessile characteristic of plants. We argue that it is essential to position plants as immobile beings in designing interactive systems. It means when seeing plants as immobile, designers can recognise the more active agency of plants and the stronger bond with which plants constantly intertwine with and manipulate their surroundings [3]. Tracing back to an earlier source, the botanist Albert Seward [1932] even suggested that such immobility constraints indicate a kind of plant-specific environmental intelligence. And even humans are not excluded from plants' manipulative behaviours. Following these theoretical perspectives from biology, we took serious consideration of the sessile activeness and silent intelligence of plants in the design of our interactive system 'Slooooooooow'.

Based on the fundamental concept of plants as immobile, this perspective requires a holistic approach focusing on the dynamic relationships among actors. In our experiment, we positioned both humans and plants as two embedded elements in the whole microclimate environment where they both influence each other. It is demonstrated in our strategy of choosing the humidity level as the input feature of the installation. From the perspective of microclimate, the humidity level results from multiple players. On the one hand, plants play an important role in shaping the humidity level of the microclimate, yet it is not exclusive. On the other hand, the humidity level can be altered by other local elements, such as wind or the presence of human users.

Therefore, as a practical suggestion for human-plant interaction, we suggest using an inclusive feature of the microclimate, such as the humidity level, to deliver the complexity and subtlety of locality to the digital world. Similarly, another example of an inclusive feature is the sunlight intensity that includes the locality's complexity, such as each floating cloud and shadow of all kinds of physical objects. Inclusive features are often environmental features, thus, allowing the participation of all local agents, including the often-neglected ones. For instance, in our experiment, the humidity sensor's position, close to the three plants or in the shade, shapes the humidity level. Compared with the changes in plant voltages due to the human acts of touching or watering, the changes in the microclimate are caused by the human presence through each inhale and exhale or through sweating. Thus, it reveals a more subtle, continuous, and bidirectional relationship between humans and plants. Therefore, an inclusive feature transforms the human-plant relationships from objectification to unification.

## 5 CONCLUSION

How can we connect the hybrid elements of humans, plants and technological devices interactively? How can we sensitise humans to the movements of non-human entities such as plants or the microclimate? We have shown an experiment of the IoT-based interactive installation 'Slooooooooooow'. The system associated the biological slowness of plants with the technological slowness of IoT homes. Thus, our experiment opens the design space of designing for a slow domestic technological environment based on the biological features of plants, humans or the microclimate as a whole. In this way, the technological home becomes the new hybrid body with the merges of Biosphere and Technosphere. It echoes Hallnäs & Redström's argument [2001] of designing for the 'structures within which we express the presence and build our "work-worlds" and "life-worlds" through interaction with the environment.'

As the initial result of this design experiment, we propose a new perspective called 'plant-decentred perspective' when designing for human-plant interaction. This plant-decentred perspective is illustrated in our strategy of choosing the humidity level as the input for eliciting interaction between technological devices and plants. The humidity level, considered the inclusive feature, is the result of the interaction and participation of all actors in the microclimate of the environment. This strategy shows our relationship-oriented perspective instead of the object-oriented one.

Furthermore, this experiment reminds us how much we can take the human-centred perspective for granted. Just like positioning a human as an individual entity, we would see a plant as an entity, and the interaction would be about the physical touch of the human finger on the plant. In this way, we are isolating the plant from the surroundings by focusing on, for instance, how it feels instead of how it influences or is influenced by others. Our work rejects the objectification and isolation of plants. Meanwhile, it suggests a new realm of human interventions beyond vision and hearing, such as the breath. It reinforces that every element is all part of this microclimate through non-human and human metabolisms. Therefore, as the larger implication, this study helps humans experience and interact with plants in a more empathic way. Equally important, it reflects the living condition of humans ourselves. How do we see ourselves as humans and our relationships with others? Our experiment shows if we experience and influence the room from the perspective of the humidity, there is no difference between us or others, following what Kirksey & Helmreich [2010, p. 562-563] claim:

'The goal in multispecies ethnography should not just be to give voice, agency or subjectivity to the non-human—to recognise them as others, visible in their difference—but to force us to radically rethink these categories of our analysis as they pertain to all beings.'

# REFERENCES

[1] Ofer Asaf, Ronnie Oren, Shachar Geiger, and Michal Rinott. 2021. Plantimus–A Plant Stethoscope. In *Proceedings of the Fifteenth International Conference on Tangible, Embedded, and Embodied Interaction*, 1-5.

[2] Fredrik Aspling, Jinyi Wang, and Oskar Juhlin. 2016. Plant-computer interaction, beauty and dissemination. In *Proceedings of the Third International Conference on Animal-Computer Interaction*, 1-10.

[3] František Baluška and Stefano Mancuso. 2020. Plants, climate and humans: plant intelligence changes everything. *EMBO reports 21*, 3, e50109.

[4] Andrea Botero. 2021. Julia Watson: Lo-TEK. Design by Radical Indigenism. *Tecnoscienza-Italian journal of science & technology studies 12*, 1, 148-151.

[5] Daniel Chamovich. 2012. *What a Plant Knows: A Field Guide to the Senses*. Scientific American/Farrar, Straus and Girou.

[6] Mads Bering Christiansen, Jonas Jørgensen, Anne-Sofie Emilie Belling, and Laura Beloff. 2020. Soft robotics and posthuman entities. *Journal for Artistic Research 22*.

[7] Batya Friedman and Lisa P Nathan. 2010. Multi-lifespan information system design: A research initiative for the HCI community. In *Proceedings of the SIGCHI Conference on Human Factors in Computing Systems*, 2243-2246.

[8] Jörg Fromm and Silke Lautner. 2007. Electrical signals and their physiological significance in plants. *Plant, cell & environment 30*, 3, 249-257.

[9] Peter Haff. 2014. Humans and technology in the Anthropocene: Six rules. *The Anthropocene Review 1*, 2, 126-136.

[10] Lars Hallnäs and Johan Redström. 2001. Slow technology–designing for reflection. *Personal and Ubiquitous Computing 5*, 3, 201-212.

[11] S Eben Kirksey and Stefan Helmreich. 2010. The emergence of multispecies ethnography. *Cultural anthropology 25*, 4, 545-576.

[12] Dov Koller and Elizabeth Van Volkenburgh. 2011. *The restless plant*. Harvard University Press.

[13] Ann Light, Irina Shklovski, and Alison Powell. 2017. Design for existential crisis. In *Proceedings of the 2017 CHI Conference Extended Abstracts on Human Factors in Computing Systems*, 722-734.

[14] Jen Liu, Daragh Byrne, and Laura Devendorf. 2018. Design for collaborative survival: An inquiry into human-fungi relationships. In *Proceedings of the 2018 CHI Conference on Human Factors in Computing Systems*, 1-13.

[15] Clara Mancini and Jussi Lehtonen. 2018. The emerging nature of participation in multispeciess interaction design. In *Proceedings of the 2018 Designing Interactive Systems Conference*, 907-918.

[16] S. Mancuso. 2018. *The revolutionary genius of plants: a new understanding of plant intelligence and behavior*. Simon and Schuster.

[17] William Odom, Richard Banks, Abigail Durrant, David Kirk, and James Pierce. 2012. Slow technology: critical reflection and future directions. In *Proceedings of the Designing Interactive Systems Conference*, 816-817.

[18] Michael Pollan. 2002. *The botany of desire: A plant's-eye view of the world*. Random house trade paperbacks.

[19] Ivan Poupyrev, Philipp Schoessler, Jonas Loh, and Munehiko Sato. 2012. Botanicus Interacticus: interactive plants technology. In *ACM SIGGRAPH 2012 Emerging Technologies*, 1-1.

[20] Elena Sabinson, Isha Pradhan, and Keith Evan Green. 2021. Plant-Human Embodied Biofeedback (pheB): A Soft Robotic Surface for Emotion Regulation in Confined Physical Space. In *Proceedings of the Fifteenth International Conference on Tangible, Embedded, and Embodied Interaction*, 1-14.

[21] Jinsil Hwaryoung Seo, Annie Sungkajun, and Jinkyo Suh. 2015. Touchology: towards interactive plant design for children with autism and older adults in senior housing. In *Proceedings of the 33rd Annual ACM Conference Extended Abstracts on Human Factors in Computing Systems*, 893-898.

[22] Albert Seward. 1932. *Plants what They are and what They Do*. CUP Archive.

[23] Percy Smith. 1910. The Birth Of A Flower Retrieved May 30, 2022 from https://www.youtube.com/watch?v=C5eAEXKJRmA.

[24] Mischer' Traxler Studio. 2021. The idea of a tree Retrieve from https://mischertraxler.com/projects/the-idea-of-a-tree-process/.

[25] Bbc Studios Natural History Unit. 2022. The Green Planet Retrieved from https://www.bbcearth.com/shows/the-green-planet.